\documentclass[aps, reprint, superscriptaddress]{revtex4-2}
\usepackage{newtxtext, newtxmath}
\usepackage{dcolumn}
\usepackage{braket}
\usepackage{graphicx}
\usepackage{orcidlink}
\usepackage{siunitx}
\usepackage{color}
\definecolor{bblue}{rgb}{0, 0.0, 0.8}
\definecolor{rred}{rgb}{0.7, 0.0, 0.0}

\usepackage{hyperref}
\hypersetup{
    colorlinks=true,        
    linkcolor=rred,          
    citecolor=rred,        
    filecolor=rred,      
    urlcolor=rred           
}

\usepackage{dcolumn}

\begin{document}
\title{Non-linear diffusion and inhomogeneity of the magnetic field in single-turn coils: Insights from 3D multiphysics modeling}
\author{Hideaki Kobayashi}
\author{Yugaku Goyo}
\affiliation{Department of Engineering Science, University of Electro-Communications, Chofu, Tokyo 182-8585, Japan}
\author{Yuto Ishii\orcidlink{0000-0001-6849-8915}}
\author{Yasuhiro H. Matsuda\orcidlink{0000-0002-7450-0791}}
\affiliation{Institute for Solid State Physics, Kashiwa, Chiba 277-8581, Japan}
\author{Kunio Takekoshi}
\affiliation{Terrabyte Co., Ltd, Yushima, Bunkyo-ku, Tokyo, 113-0034, Japan}
\author{Akihiko Ikeda\orcidlink{0000-0001-7642-0042}}
\email[Contact author: ]{a-ikeda@uec.ac.jp}
\affiliation{Department of Engineering Science, University of Electro-Communications, Chofu, Tokyo 182-8585, Japan}

\date{\today}

\begin{abstract}
The single-turn coil method is a destructive pulsed magnet for generating over 100 T with a few $\mu$-second pulse duration, and it inevitably causes the coil to explode.
The temporal and spatial distributions of the electric current and magnetic field are highly inhomogeneous, arising from the skin effect, rapid temperature rise, and coil deformation.
To grasp the dynamic phenomena in the single-turn coil, we conducted a finite element analysis using multiphysics simulation.
We employed finite element method calculations using a fully 3D model of the single-turn coil with broken cylindrical symmetry.
The calculated result revealed highly nonlinear diffusion of electric current, temperature, and magnetic fields, which are the sources of the inhomogeneous magnetic fields inside the single-turn coil in time and space.
\end{abstract}

\maketitle

\section{Introduction}
The ultrahigh magnetic field region beyond 100 T, extending to 1000 T, is an experimental frontier of science.
Recently, novel phase transitions have been discovered in the 1000 T region in such condensed matters as solid O$_{2}$ \cite{NomuraPRL2014, NomuraOxygen2022} , CdCr$_{2}$O$_{4}$ \cite{MiyataPRL2011, GenPRL2026}, VO$_{2}$ \cite{MatsudaNC2020, MatsudaJJAPProc2026}, FeSi \cite{NakamuraPRL2021}, YbB$_{12}$\cite{TerashimaJPSJ2018}, SmB$_{6}$ \cite{NakamuraPRB2022}, LaCoO$_{3}$ \cite{IkedaPRB2016,  IkedaPRL2020,  IkedaNC2023}, SrCu$_2$(BO$_3$)$_2$ \cite{MatsudaPRL2013, NomuraNC2023}.
These discoveries are made using improved techniques of the single turn coil method and the electro-magnetic flux compression method, and also using the new measurement technique ranging from magneto-optical effects \cite{MiyataPRL2011}, the high-frequency conductivity \cite{KohamaRSI2020, ShitaokoshiRSI2023, PengSST2025}, magnetization \cite{TakeyamaJPSJ2012, GenPNAS2023}, magnetostriction \cite{RodriguezOE2015, IkedaRSI2017}, ultrasound \cite{NomuraRSI2021}, dielectric constant \cite{ChiuJAP2025}
Modification of the single-turn coil method, such as putting crey on the outside of the single-turn coil
\cite{GenRSI2021}, and putting a slit by removing the center volume of the single turn coil \cite{, IshiiJJAPProc2026} are reported.

When generating an intense magnetic field above 100 T, the coil is subjected to tensile stress of 4 GPa, leading to the coil's explosion.
This feature makes the research area beyond 100 T an experimental frontier.
The explosion is the consequence of the energy density of the magnetic field expressed as $E/V = B^{2}/(2\mu_{0})$ whose unit is [J/m$^{3}$], which is equal to [Pa] the pressure. 
It states that confining a magnetic field of 100 T and 1000 T inevitably requires 4 GPa and 400 GPa, respectively.
That pressure is applied to the coil, causing it to explode.
It also states that one definitely needs a high pressure to achieve well beyond 100 T and to reach 1000 T, which is realized by the implosion of the liner in the flux compression methods \cite{HerlachRPP1999, NakamuraRSI2018}

The aforementioned methods are achieved by overcoming or avoiding the characteristics of the single-turn coil, the short $\mu$-second pulse, explosion, electromagnetic noise, single-shot nature, and highly inhomogeneous magnetic field in time and space.
Recently, notable efforts have been made in the above-100 T region, ranging from new techniques to make the single-turn coil portable \cite{IkedaAPL2022}.
The technique has enabled the x-ray experiment using x-ray free electron laser \cite{IkedaPRL2025}, which opens a new era for the 100 T science.

Single turn coil has been devised experimentally by Herlach \cite{HerlachIEEE1971, HerlachJPE1973, HerlachPRB1974}, in Tokyo \cite{NakaoJPE1985, TakeyamaJPE1988, SakakibaraJPCM1990, KatoriJPSJ1995}, and in Berlin \cite{PortugallJPD1997, PuhlmannJPD1997}.
See the reviews \cite{HerlachRPP1968, HerlachRPP1999}.
Previously, finite-element methods has been employed to simulate the flux-compression methods \cite{MiuraJJAP1990, NakamuraRSI2014} and the single-turn coil methods \cite{MiuraJJAP1990, GeJAP2023, GePF2024a, GePF2024b, GePF2024c, GePS2024a, GePS2024b, GeJAP2024, GePS2025a, GePS2025b}.
All these studies assume the cylindrical symmetry of the model.
However, in considering the inhomogeneity of the single-turn coil, it is crucial to take into account the feed gap (neck part) of the single turn coil.

Here, we employed finite element method calculations using a fully 3D model of the single-turn coil with a broken cylindrical symmetry.
The calculated result revealed a non-linear diffusion of electric current, temperature, and magnetic fields, which are the source of the inhomogeneous magnetic fields inside the single-turn coil in time and space.

\section{Method}

\begin{figure*}
\begin{center}
\includegraphics[width = \textwidth]{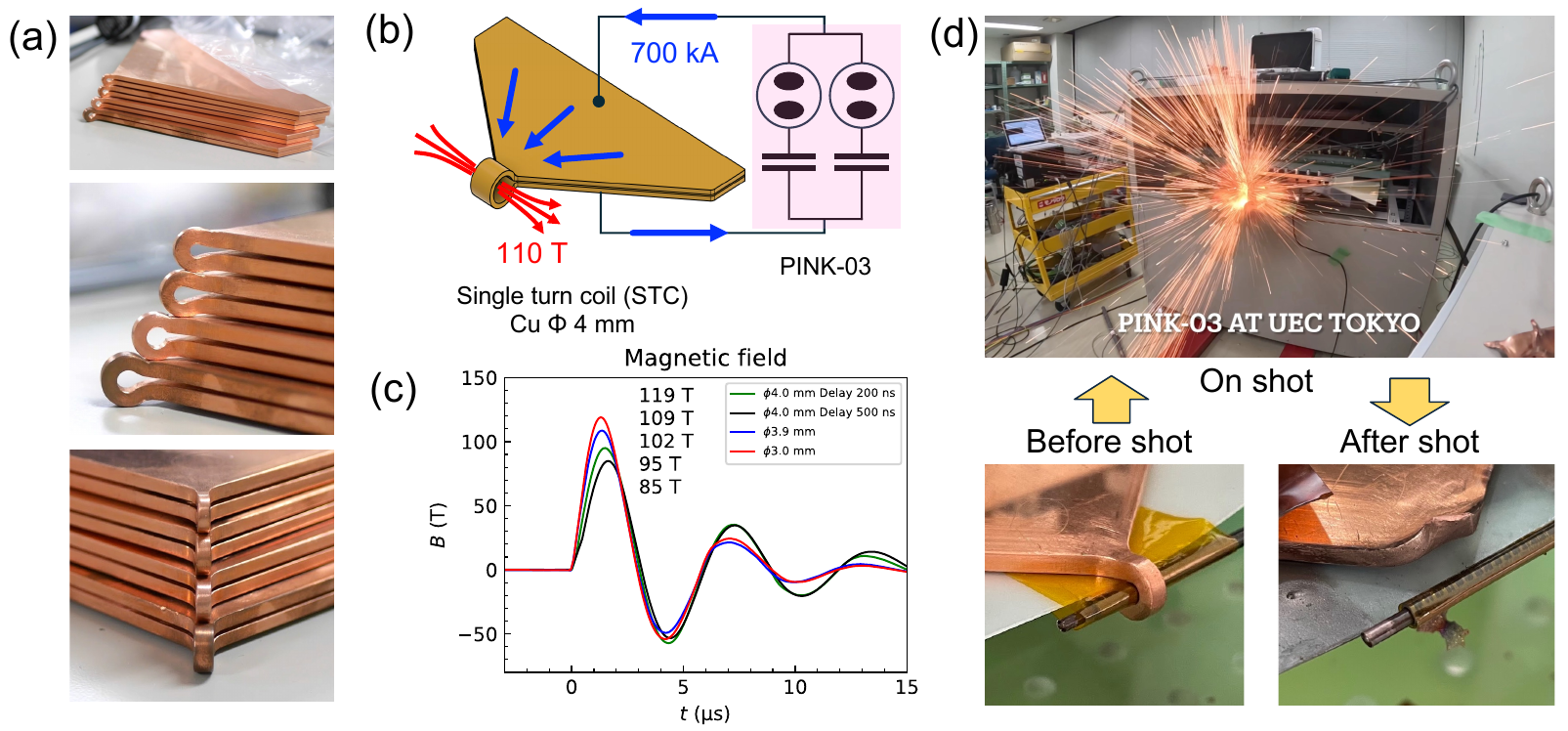}
\caption{(a) The photos of the single turn coils of $\phi$3 to $\phi$4 mm.
(b) The schematic drawing of the electric circuit.
(c) The obtained magnetic field waveform.
(c) The photos of the before, during, and after the 120 T generation.
\label{pink03}}
\end{center}
\end{figure*}

Experimentally, we generated a 120 T using the PINK-03 in the University of Electro-Communications, Tokyo, Japan.
PINK-03 is an identical system to PINK-02, devised for the x-ray free electron laser experiment in Ref. \cite{IkedaPRL2025}.
Fig. \ref{pink03} shows the photos of the single turn coils of $\phi$3 to $\phi$4 mm, the schematic drawing of the electric circuit, the obtained magnetic field waveform, and the photos of the before, on, and after the 120 T generation.

\begin{figure*}
\begin{center}
\includegraphics[width = \textwidth]{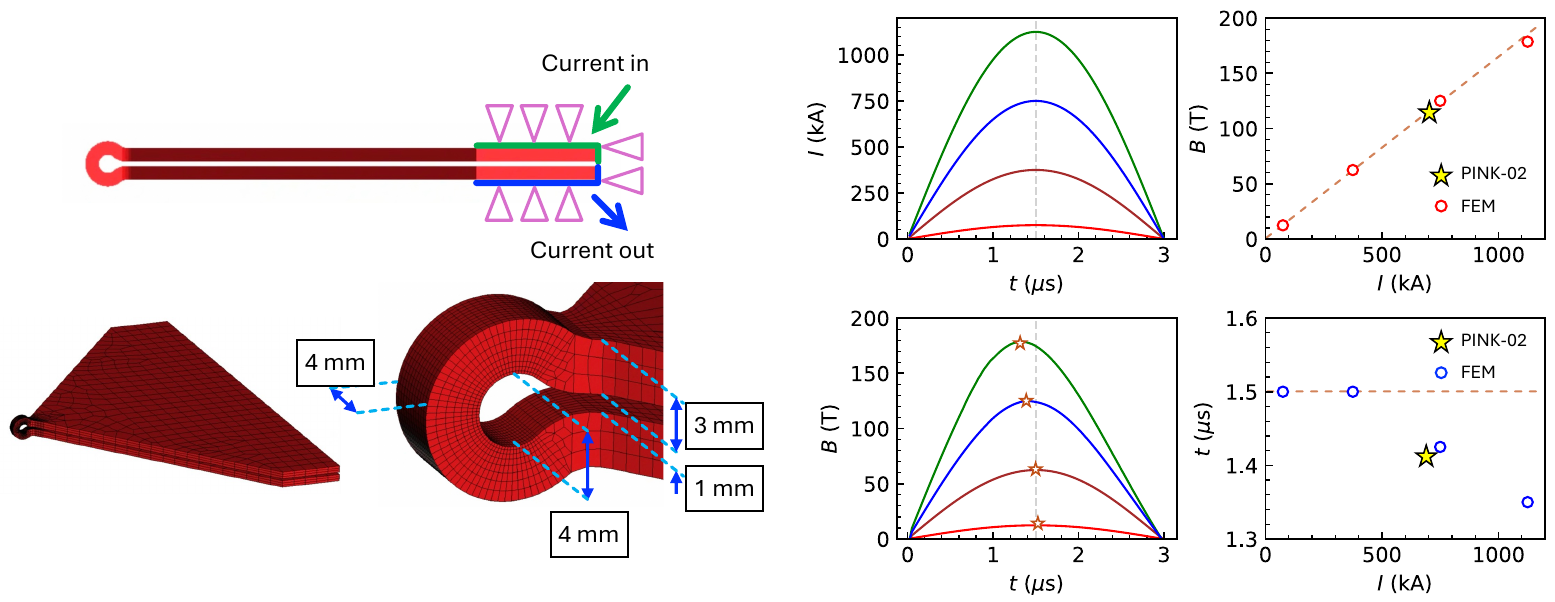}
\caption{The finite element model of a single turn coil of the diameter $\phi4$ mm.
The profiles of the input current and the obtained magnetic fields at the center position of the single-turn coil.
The relation between the maximum current and the maximum magnetic field at the center position.
The relation between the maximum current and the time to reach the maximum magnetic field at the center position of the single-turn coil.
\label{model}}
\end{center}
\end{figure*}

\begin{figure*}
\begin{center}
\includegraphics[width = \textwidth]{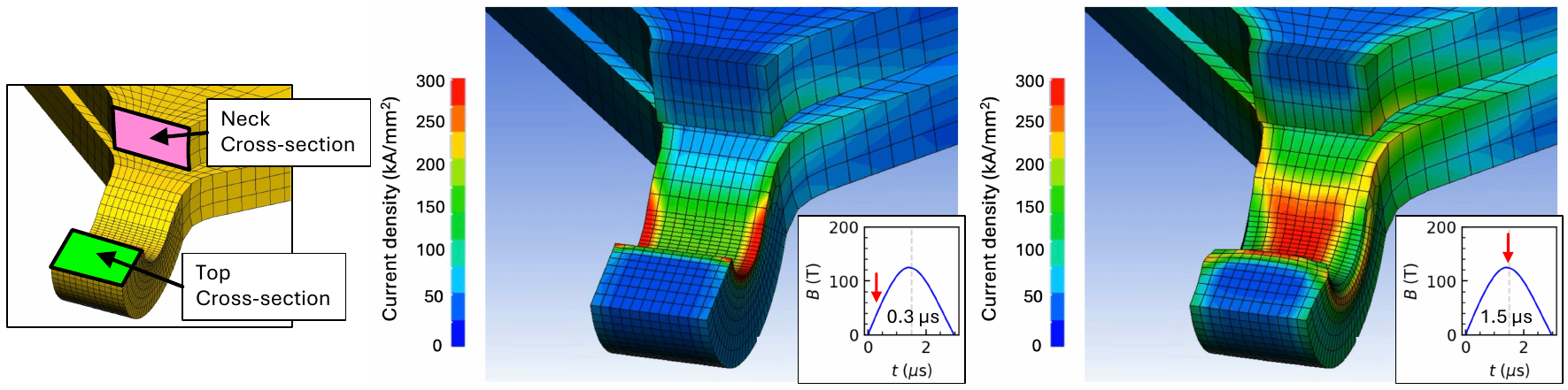}
\caption{The representative result of the calculation in the 120 T generation using the 750 kA current.
The cross-sectional view of the neck part and the top part of the single turn coil at 0.3 $\mu$s and at 1.5 $\mu$s, which are generating 50 T (an early stage) and 120 T (the maximum stage), respectively.
\label{result1}}
\end{center}
\end{figure*}

The single-turn coil of $\phi 4$ mm is modeled using {\footnotesize FUSION360}.
The finite element model is using {\footnotesize  Altair HyperWorks} with ??? elements as shown in Fig. \ref{model}.
To account for the skin effect, smaller elements are placed in the proximity of the inner surface of the coil.
{\footnotesize Ansys LS-DYNA Ver. R13.1.1} is used for the multiphysics FEM calculation.
A Threadripper PRO 5995WX (64-core CPU) with 512 GB of random access memory is used.
The obtained results are rendered and analyzed using {\footnotesize LS-Prepost}.
The current waveform of 3 $\mu$s duration and a peak of 750 kA is employed from Ref. \cite{IkedaAPL2022, IkedaPRL2025}, which is scaled to 75, 375, 750, and 1125 kA as shown in Fig. \ref{model}.
The obtained magnetic field profile is shown in Fig. \ref{model}.
Fig. \ref{model} shows the relation between the maximum current and the maximum magnetic field at the center position, where a slight decrease from the linear extrapolation is observed above the 1125 kA case.
Fig. \ref{model} shows the relation between the maximum current and the time to reach the maximum magnetic field at the center position of the single-turn coil, where the forward shift of the time to reach the maximum magnetic field is observed in cases of 750 kA and 1125 kA.
These results are quite consistent with the experimental observations, indicating that the series of calculations was conducted properly.

In the multiphysics finite element method calculations, the mechanical, electromagnetic, thermal effects, and the transient change of material properties are taken into account, where the dynamical electric current flow, the generation of the magnetic field, the mechanical deformation, the resistive heating, the temperature change, the resistivity change are involved. 
The magnetic field distribution is measured at various positions around the coil.

\section{Result and Discussion}

\subsection{Representative result}

Representative result of the calculation for 120 T generation at 750 kA current.
The cross-sectional view of the neck part and the top part of the single-turn coil are shown in Fig. \ref{result1}.
The clear difference in the electric current distribution is observed between the 0.3 $\mu$s panel and the 1.5 $\mu$s, which are the early stage (50 T) and the max stage (120 T) of the magnetic field distribution.
In the 0.3 $\mu$s panel, the electric current is predominantly concentrated at the two edges of the inner surface of the single-turn coil.
On the other hand, in the 1.5 $\mu$s panel, the current is moved away from the edge part and more concentrated at the middle part of the inner surface of the single turn coil.

The electric current at the edge of the copper single-turn coil due to the skin effect, which originates in the electromagnetic induction and is represented by the skin depth using the electric resistivity and the current frequency as $\delta = \sqrt{\rho/(\pi f \mu)}$.
In the present case, we obtain $\delta = 0.160$ mm for copper $\rho = 1.69\times10^{-8}$ \unit{\ohm}$\cdot$m, $\mu\simeq4\pi\times10^{-7}$, and the $f = 1.66 \times 10^{5}$, which is in agreement with the 0.3 $\mu$s panel of Fig. \ref{result1}.
The observed time-dependent change in the path of the electric current within the cross-section of the single-turn coil is the highlight of the present study.
This dynamic diffusion of the electric current is the source of the highly non-uniform, time-dependent distribution and changes in the generated magnetic field.
The distribution of the current is not directly measurable in the experiment, making the calculation a valuable approach.

\begin{figure*}
\begin{center}
\includegraphics[width = \textwidth]{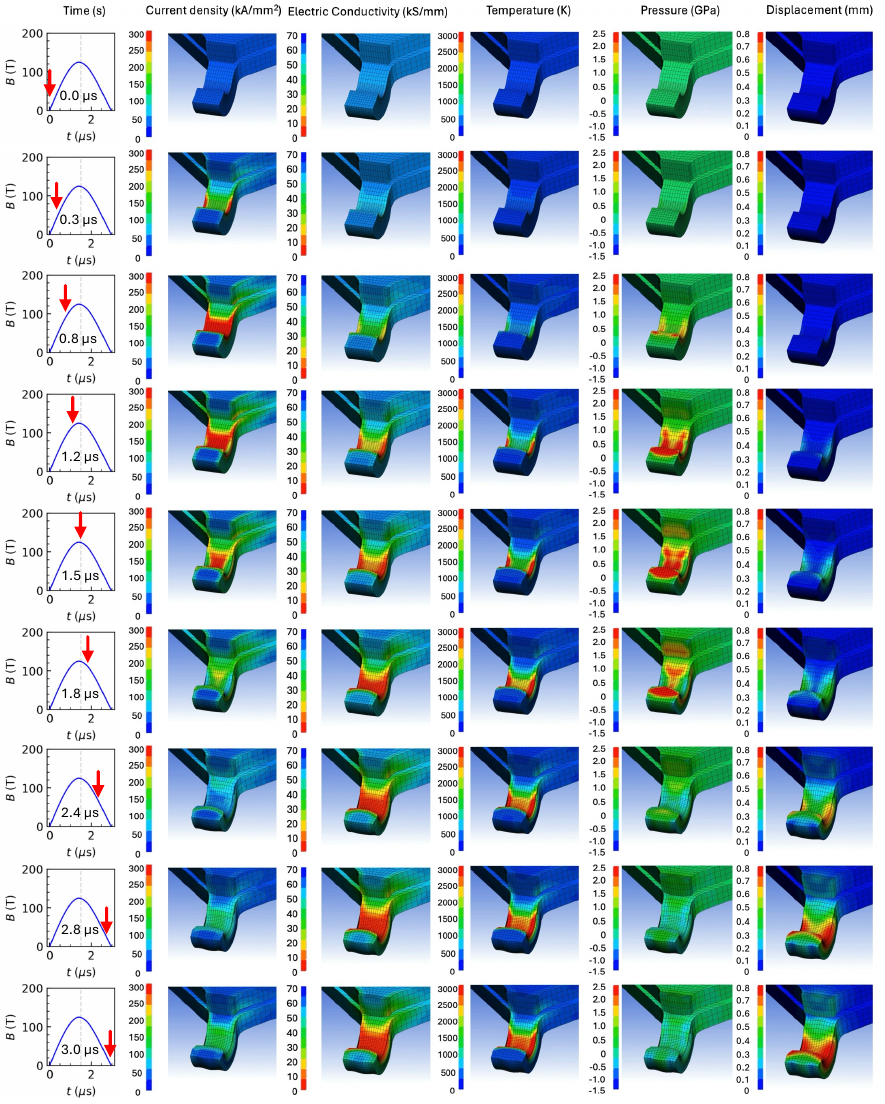}
\caption{The time evolution of the current distribution, electrical conductivity, temperature, pressure, and displacement of each finite element as shown in the cross-sectional view.
\label{result_all}}
\end{center}
\end{figure*}

\subsection{Time and position dependence of the physical properties of the single turn coil}

The time-dependent electric current distribution in the single-turn coil can be understood in terms of the time-dependent temperature rise at the edge of the single-turn coil, which increases the electric resistivity.
The scenario is evident by observing the time evolution of the current distribution, electrical conductivity, temperature, pressure, and displacement of each finite element, as shown in the cross-sectional view in Fig. \ref{result_all}.

First, we look at the time-dependence of the electric current density in Fig. \ref{result_all}.
At 0.3 $\mu$s, the electric current density concentrates at the edge of the single-turn coil.
At 0.8 $\mu$s, it is gradually diffused and spread over all the inner surface of the single-turn coil.
At 1.2 $\mu$s, the depletion of the electric current density at the edge of the single-turn coil is obvious.
At 1.5 $\mu$s, which is close to the maximum magnetic field time, the electric current density is more concentrated in the middle region of the inner surface of the single-turn coil.
At 1.8 $\mu$s and later times, the electric current density is further diffused from the surface to the interior of the single-turn coil.

Next, we look at electric conductivity, temperature, pressure and displacement in Fig. \ref{result_all}.
At 1.2 $\mu$s, it is observed that the electric conductivity has decreased at the edge of the single-turn coil.
This originates from the significant increase in temperature, which results from Joule heating due to the concentrated electric current density.
At 0.8 $\mu$s, one should note that the pressure beyond 2 GPa has been applied to the edge of the single-turn coil, which indirectly points to the location of the electric current density.
However, the displacement has not yet started.
Even at 1.2 $\mu$s, the displacement is not significant yet.
This observation claims that the diffusion of the electric current density is driven by the increase in the electric resistivity due to the temperature rise, and not by the deformation of the coil that would result in the increase in inductance.

\begin{figure*}
\begin{center}
\includegraphics[width = \textwidth]{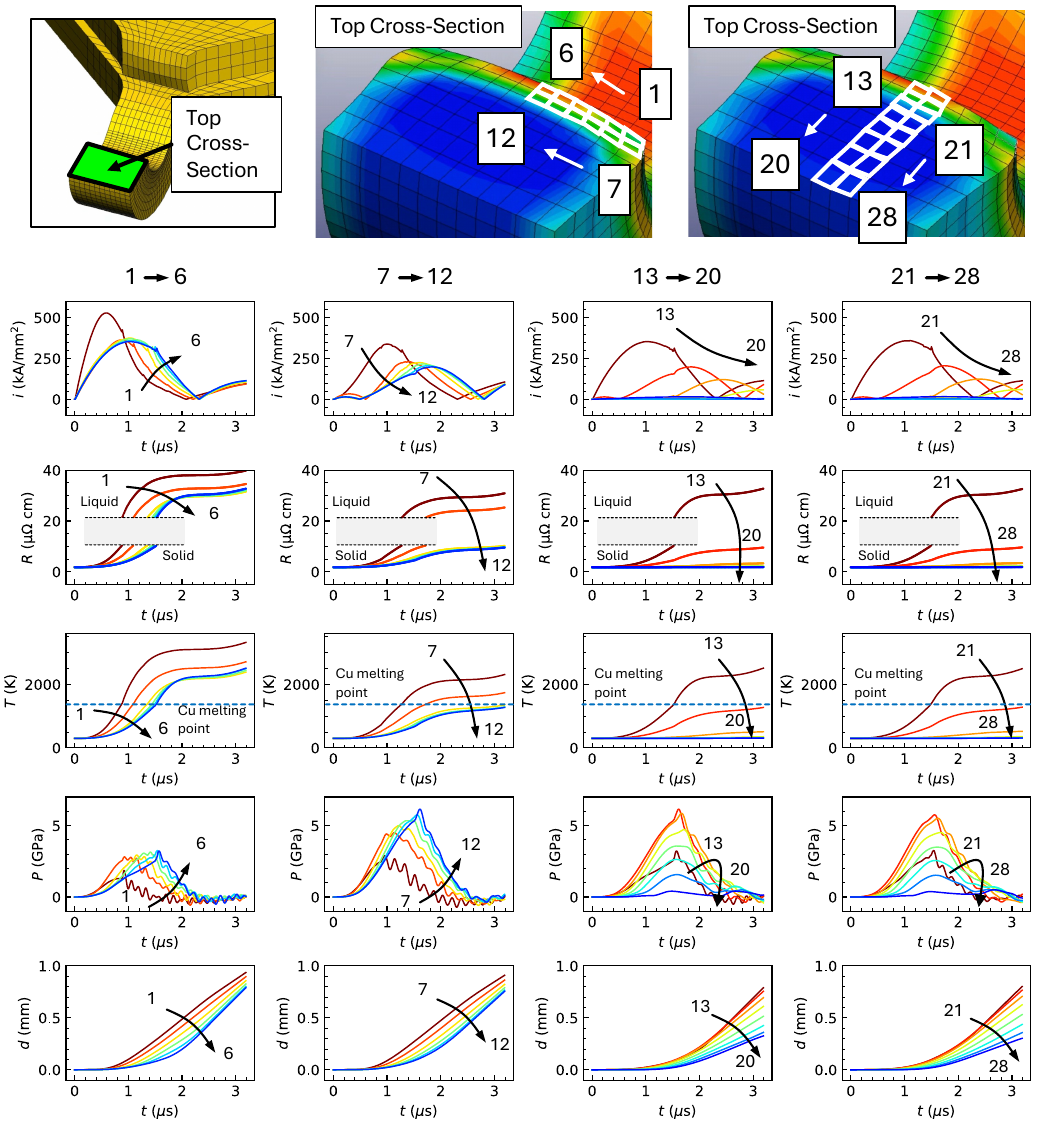}
\caption{Quantitative time-evolution of electric current density, resistivity, temperature, pressure, and displacement measured at representative elements in the top-cross-section.
\label{results2}}
\end{center}
\end{figure*}

\begin{figure*}
\begin{center}
\includegraphics[width = \textwidth]{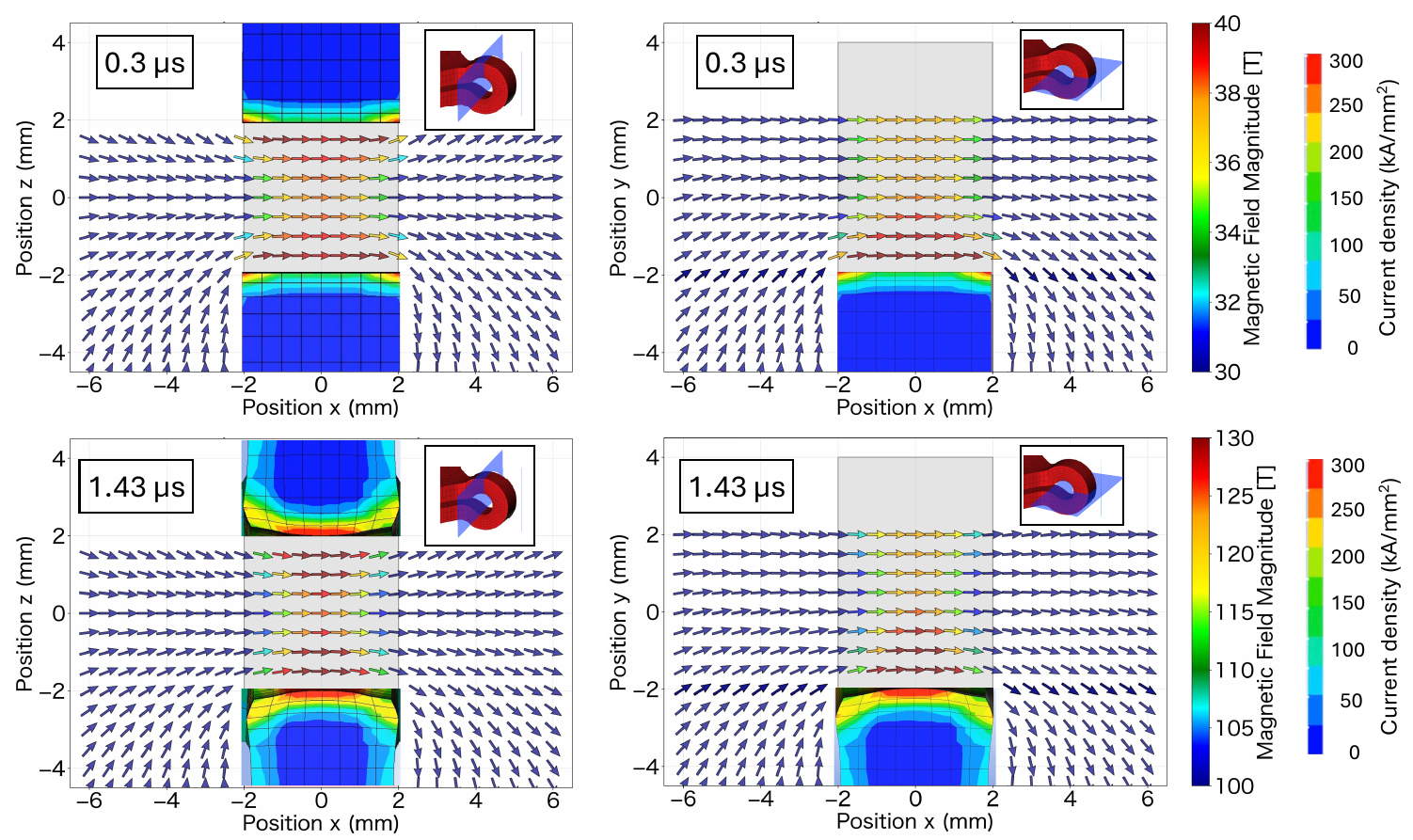}
\caption{The distribution of the generated magnetic field inside the single-turn coil at 0.3 $\mu$s and 1.43 $\mu$s.
\label{dist1}}
\end{center}
\end{figure*}

\begin{figure*}
\begin{center}
\includegraphics[width = \textwidth]{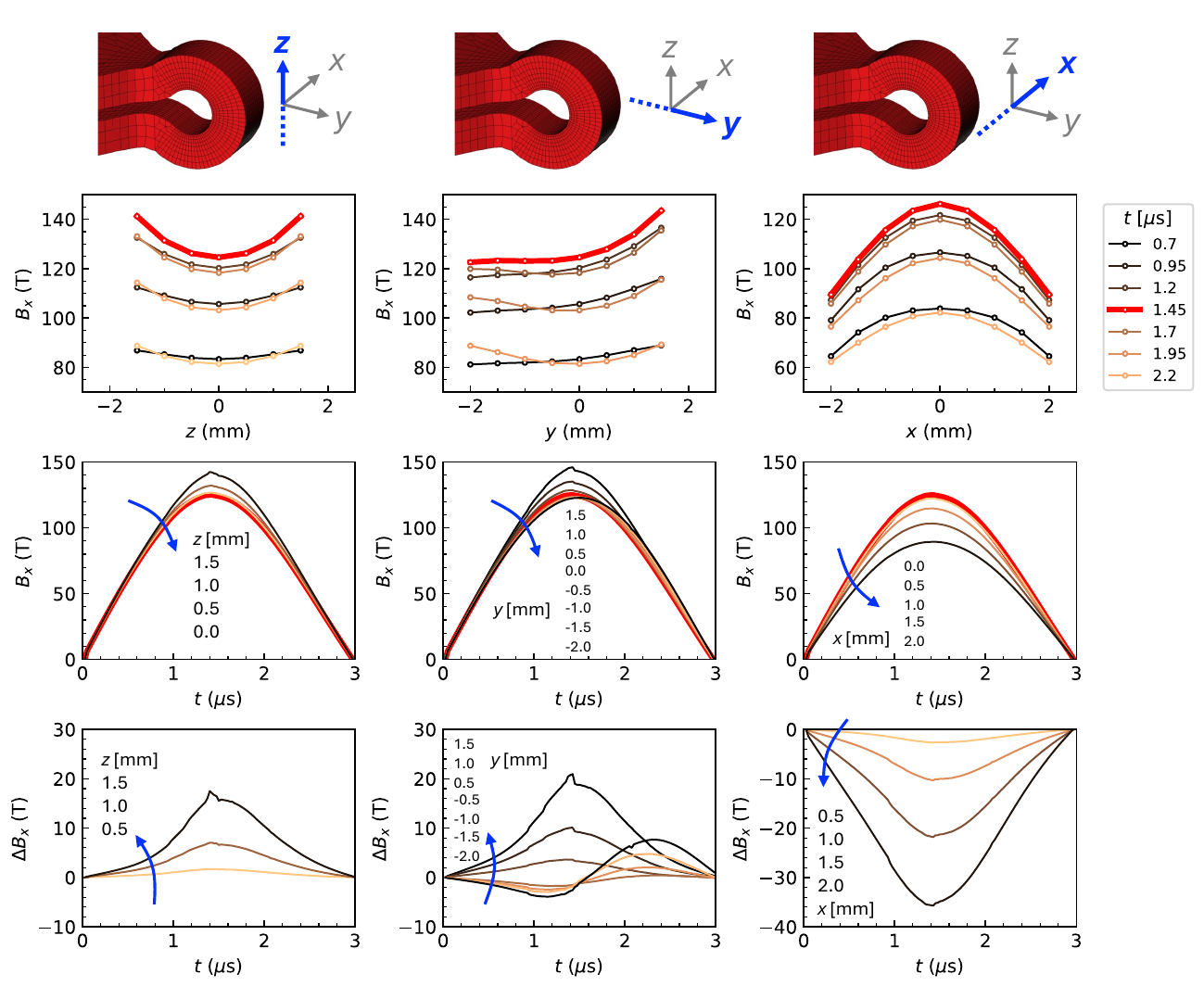}
\caption{The line profles of the spatial distribution of the generated magnetic field by the single-turn coil.
The time-dependence of the magnetic field in representative positions in the single-turn coil.
\label{dist2}}
\end{center}
\end{figure*}

We consider what happens in the latter half of the pulse.
The temperature keeps increasing at the edge and at the inner surface, leading to decreased electric conductivity at the inner surface of the single-turn coil.
This is the origin of the further diffusion of the electric current density into the interior of the copper single-turn coil.
The time-dependent location of the electric current density is also evident in the pressure profile.
At 0.8 $\mu$s, the high pressure is located at the edge of the inner surface of the single-turn coil.
At 1.2 $\mu$s, 1.5 $\mu$s, and 1.8 $\mu$s, we observe the gradual relocation of the high-pressure region from the edge to the middle part, where the two regions merge into one, which is then further diffused into the interior of the single-turn coil.

In the times after 1.8 $\mu$s, the displacement becomes evident starting from the edge region and uniformly expanding inner radius of the single-turn coil.
They do not occur simultaneously with the application of pressure due to inertia.
In the single-turn coil method, the high-pressure problem is cleverly avoided by generating the high magnetic field and conducting all experiments before the coil is destroyed.

The present discussion can be corroborated by examining the quantitative time evolution of electric current density, resistivity, temperature, pressure, and displacement measured at representative elements, as shown in Fig. \ref{results2}.
The elements 1 to 6 and 7 to 12 represent the position-dependent time evolution at the edge and middle regions of the inner surface of the single-turn coil.
The elements 13 to 20 and 21 to 28 represent the position-dependent time evolution from the inner surface to the interior of the single-turn coil.

It is reconfirmed that the electric current density is redistributed from the edge region to the middle region, as shown by the elements 1 to 6 and 7 to 12 in Fig. \ref{results2}.
Then, as evident by the elements 13 to 20 and 21 to 28, the electric current density is diffused into the interior of the copper conductor of the single-turn coil.

The resistivity in Fig. \ref{results2} shows the rapid increase at the edge region followed by the middle region.
It is noted that a discontinuous increase of the resistivity is observed in not all but multiple Elements, which originates in the melting of copper.
This is confirmed by observing the time-dependent temperature of each element, with the melting temperature reached at the inner surface and the edge region.

The pressure exceeds 1 GPa in many elements, especially in the time region between 1 $\mu$s and 2 $\mu$s, which well exceeds the failure pressure of copper, leading to plastic deformation.
As mentioned, the displacement occurs at 1 $\mu$s and becomes significant only after 2 $\mu$s, which ensures the 100 T generation and stable experiment in the single turn coil method.
Note that due to the deformation of the coil, the magnetic field profiles are quite different between the former and the latter halves of the 100 T pulse.

\subsection{Time dependence of the inhomogeneity of the magnetic field}

We observe the distribution of the generated magnetic field inside the single-turn coil as shown in Fig. \ref{dist1}.
In the vertical cross-sectional panels at 0.3 $\mu$s and at 1.43 $\mu$s, the magnetic field show a saddle point like distribution with regard to the coil center, where the magnetic field increases when approaching the inner surface of the single-turn coil and the magnetic field decreases when $x$ is increased or decreased both at 0.3 $\mu$s and at 1.43 $\mu$s.
It is interesing that one can obtain even higher magnetic field in the proximity of the coil surface than the center of the coil, as is reported in Ref. \cite{TakeyamaJPSJ2012}.
This is more evidently shown in the $z$-cut of Fig. \ref{dist2}. 

When we compare the magnetic field distribution between 0.3 $\mu$s and 1.43 $\mu$s, one notices that the distribusion is less severe in 0.3 $\mu$s than that of 1.43 $\mu$s.
This is because the electric current distribution is concentrated at the two edges of the inner surface of the single-turn coil 0.3 $\mu$s that is similar to the Helmholtz coil geometry that is known to host a homogeneous magnetic field distribusion in its axial direction.
While, it is concentrated in the middle region at 1.43 $\mu$s.
This is more evidently shown in the $x$-cut and $z$-cut of Fig. \ref{dist2}, where the distribution of the magnetic field is much smoother in the former half of the pulse ($<1.5$ $\mu$s) than those of the latter half of the pulse.

Next, we observe the horizontal cross-sectional panels in Fig. \ref{dist1} at 0.3 $\mu$s and at 1.43 $\mu$s.
Both panels show increaseing magnetic field when approaching the inner surface of the single turn coil.
In the proximity of the inner surface of the single turn coil, the magnetic field distribution is smoother at 0.3 $\mu$s than that of 1.43 $\mu$s.
This is again due to the concentration of the electric current density at the two edges 0.3 $\mu$s.
Quantitatively, the distribution is shown in the $y$-cut of Fig. \ref{dist2}.
Note again that the magnetic field distributions are distinct between the former half of the pulse ($<1.5$ $\mu$s) than those of the latter half of the pulse, which originates in the distinct electric current density distribution and deformation.

Finally, we look at the time-depence of magnetic field at representative positions in the single-turn coil as shown in Fig. \ref{dist2}.
In $z$-cut and $x$-cut, the magnetic field is larger and smaller when $z$ or $x$ is away from the coil center, where only the magnetic field strength is distinct from the coil center.
On the other hand, the $y$-cut data show that the time to reach the maximum magnetic field is slower when approaching the neck region from the coil center, which is in good agreements with the previous experimental report \cite{TakeyamaJPSJ2012}.
This highlights the complex interplay between the observed magnetic field and the time-dependent distribution of the electric current density in the single turn coil, especially at the neck region, which as not been explored in the previous studies using simulations.


\section{Summary}
We employed finite element method calculations using a fully 3D model of the single-turn coil with a broken cylindrical symmetry.
The calculated results revealed nonlinear diffusion of electric current, temperature, and magnetic fields.
First, the concentration of the electric current density at the edge region is due to the skin effect, which increases temperature and resistivity, leading to the diffusion of the electric current density from the edge region to the middle and interior regions of the single-turn coil.
Subsequently, due to the high pressure at the magnetic field maximum, the coil begins to deform, a phenomenon that becomes significant well after magnetic field generation.
These highly nonlinear diffusion processes of the electric current and the feed gap influence the time- and position-dependent intensity distribution of the magnetic field.
The knowledge obtained should contribute to the further development of destructive pulsed magnetic field methods.

\begin{acknowledgments}
This work is supported by JST FOREST program No. JPMJFR222W, JSPS Grant-in-Aid for Scientific Research on Innovative Areas (A) 23H04861, 23H04859, and Grant-in-Aid for Scientific Research (B) 23H01121, and MEXT LEADER program No. JPMXS0320210021.
\end{acknowledgments}

\bibliography{lsdyna_stc}


\end{document}